\documentclass[11pt]{article}
\usepackage[top=1.25in, left=1.25in, right=1.25in, bottom=1.25in]{geometry}
\usepackage{graphicx}
\usepackage{amssymb}
\usepackage{epstopdf}
\usepackage{ulem}
\usepackage{color}
\title{The packing of granular polymer chains}
\author{Ling-Nan Zou$^1$, Xiang Cheng$^1$, Mark L. Rivers$^2$, Heinrich M. Jaeger$^1$, Sidney R. Nagel$^1$}
\begin{document}

\maketitle

\footnotetext[1]{The James Franck Institute and Department of Physics, The University of Chicago, Chicago, IL 60637}
\footnotetext[2]{The Department of Geophysical Sciences and Center for Advanced Radiation Sources, The University of Chicago, Chicago, IL 60637}

\begin{abstract}
Rigid particles pack into structures, such as sand dunes on the beach, whose overall stability is determined by the average number of contacts between particles. However, when packing spatially extended objects with flexible shapes, additional concepts must be invoked to understand the stability of the resulting structure.  Here we study the disordered packing of chains constructed out of flexibly-connected hard spheres. Using X-ray tomography, we find long chains pack into a low-density structure whose mechanical rigidity is mainly provided by the backbone. On compaction, randomly-oriented, semi-rigid loops form along the chain, and the packing of chains can be understood as the jamming of these elements. Finally we uncover close similarities between the packing of chains and the glass transition in polymers.
\end{abstract}

It is an enduring puzzle why a boxful of ball bearings, even when compacted by many taps, never packs denser than $\simeq 0.64$  \cite{AsteWeaire2008, Zallen1998, Bernal1959, Scott1962, Finney1970, Knight&c1995}. This is much less dense than Kepler's stacking of triangular layers, known to every grocer as the optimal packing of oranges (density $\simeq 0.74$). Yet despite their sub-optimal density and lack of periodic order, jammed packings are rigid and resist shear. Replacing spheres with less symmetric objects, e.g., rods or ellipsoids, introduces new degrees of freedom that alters the packing structure and creates new modes of response \cite{Villarruel&c2000, WilliamsPhillipse2003, Donev&c2004, Man&c2005, Zeravcic&c2009, Mailman2009}. Here, we describe the jammed packing of flexible granular chains.  Although density and coordination number decrease dramatically with increasing chain length, the packings remain rigid.  Using X-ray tomography to visualize the chain conformations, we find long floppy chains effectively partition into a collection of nearly-rigid elements --- small loops --- that then jam into a rigid packing. Randomly-packed spheres have often been used as a model for simple glasses \cite{Zallen1998, Cahn1980}.  Building on this and on the fact that polymer molecules are often modeled as flexible chains \cite{JerniganFlory1969, CarmesinKremer1988}, we find that the jammed packing of chains captures the dependence of the polymer glass transition on chain length, topology, and stiffness.

\begin{figure}[p]
\center
\includegraphics[width=0.75\textwidth]{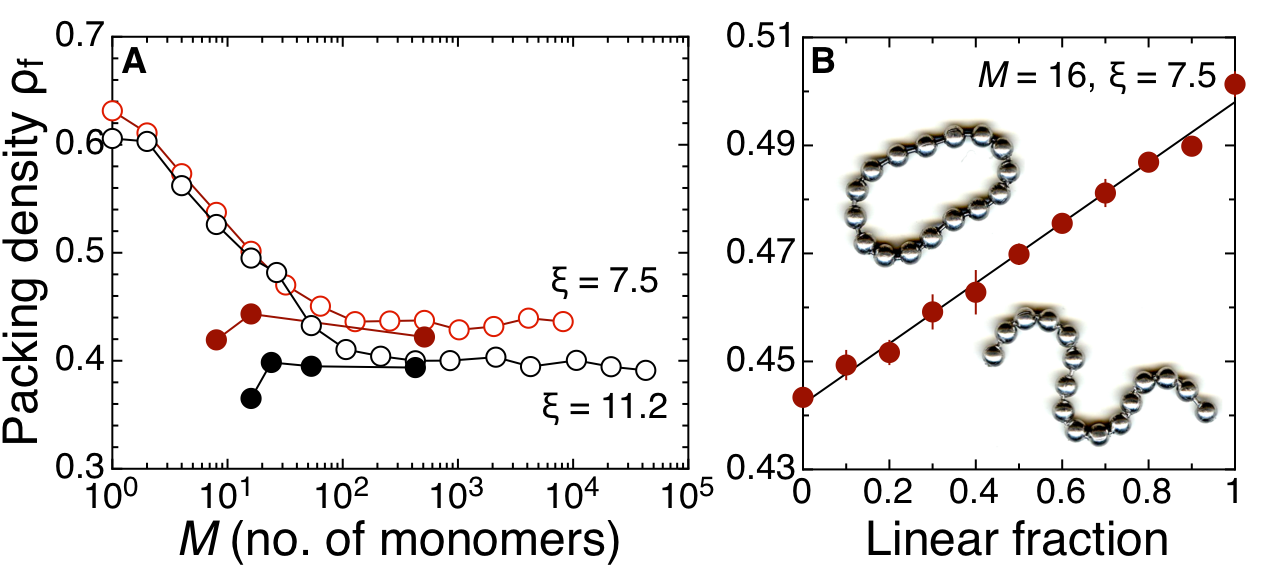}
\caption{(a) Packing density $\rho_\mathrm{f}$ of ball-chains versus chain length $M$ after $10^{4}$ taps, for both linear (open symbols) and cyclic (filled symbols) chains. Each data point is the average of 5 trials; error bars are smaller than the size of the symbols. (b) Packing density of a mixture of linear and cyclic chains (both with $M=16$) as a function of the fraction of linear chains. Solid line is a linear fit.}
\label{rho(M)}
\end{figure}

Our chains are the familiar ball-chains used as window-shade pulls (Fig.~\ref{rho(M)}b, inset): the ``monomers'' are hollow metal spheres, the ``bonds'' are short metal rods. The monomers are uniform in size and uniformly distributed along the chain; they are also free to rotate about the backbone so that the chain does not support torsion. These ball-chains are not arbitrarily flexible: there is a maximum bond flex angle $\theta_\mathrm{max}$. We use the minimum loop size $\xi=2\pi/\theta_\mathrm{max}$ as a simple measure of chain stiffness; this is the minimum length of chain that can close to form a ring. We used an aluminum chain with monomer diameter $a=2.4 \ \textrm{mm}$ and $\xi=7.5$, and a brass chain with $a = 1.9 \ \textrm{mm}$ and $\xi = 11.2$. For each, we studied the packing of both linear chains with free ends, and of cyclic chains whose ends are connected together in a loop.

The chains are loaded into a cylindrical cell mounted on an electro-mechanical shaker: long chains are rapidly unspooled into the cell end-first, short chains are poured-in few at a time. The chains are compacted by giving the cell discrete vertical ``taps'': a single period of 30 Hz sinusoidal vibration with peak-to-peak acceleration of $8 g$, where $g=9.8\ \textrm{m/s}^{2}$. After each tap, the pack height is measured to find the packing density $\rho$. The compaction process found here is similar to that of hard spheres \cite{Knight&c1995}: $\rho$ increases in a logarithmic fashion with the number of taps, and is slow to approach a steady state; we chose $10^4$ taps as an arbitrary stopping point to measure the ``final'' density $\rho_\mathrm{f}$. The packing produced is always disordered with no signs of chain crystallization. We repeated these measurements using cells of two different diameters, 4.75 cm and 2.5 cm; this had no strong effect on our results.

Fig.~\ref{rho(M)}a plots $\rho_\mathrm{f}$ versus $M$, the number of monomers per chain. For linear chains, $\rho_ \mathrm{f}$ falls monotonically with increasing $M$, from $\rho_\mathrm{f} \simeq 0.64$ for $M=1$ to a much-reduced asymptotic value $\rho_{\mathrm{f}, \infty}$ for the longest chains ($M\rightarrow \infty$). The asymptotic density is slightly higher for the floppy chain ($\xi=7.5$ and $\rho_{\mathrm{f}, \infty} = 0.43$) than for the stiff chain ($\xi=11.2$ and $\rho_{\mathrm{f}, \infty} = 0.39$). For linear chains, varying $M$ changes the density of chain ends in the pack.  For cyclic chains, end-effects are absent.  As expected, long cyclic and linear chains pack at the same asymptotic density; however, short (but still floppy) cyclic chains ($M \gtrsim 2\xi$) also packs at $\rho_{\mathrm{f}, \infty}$. Still smaller $(M \rightarrow \xi)$, semi-rigid cyclic chains actually pack less densely than the long chain limit. The trend here is opposite to that found in linear chains, where short chains pack denser than long ones. 

That small, floppy cyclic chains pack at $\rho_{\mathrm{f},\infty}$ suggests end-effects are responsible for the enhanced packing density of short linear chains (in the floppy regime). We can check this directly by packing a mixture of linear and cyclic chains of the same length. As shown in Fig.~\ref{rho(M)}b, $\rho_\mathrm{f}$ increases linearly with the fraction of linear chains.  This indicates that in a packing of floppy chains, ends act as non-interacting, density-enhancing defects.

To reveal the detailed packing structure, we used X-ray tomography. Here, only the $\xi=7.5$ Al chain is sufficiently X-ray transparent to be imaged. Our X-ray source is a 37 keV beam (GSECARS beam line) at the Argonne Advanced Photon Source. Beam size limits us to a relatively small sample: the chains are confined inside a cylinder of diameter $2.5\ \textrm{cm}$ ($=10.5$ monomer diameters), and the imaging volume covers 5 cm (out of $\sim20$ cm) of the packing column, containing 1000-1500 monomers. On the other hand, the high image resolution (27.5 $\mu\textrm{m}$/voxel) allows us accurately locate every monomer and trace every bond.

\begin{figure*}[p]
\center
\includegraphics[width=0.75\textwidth]{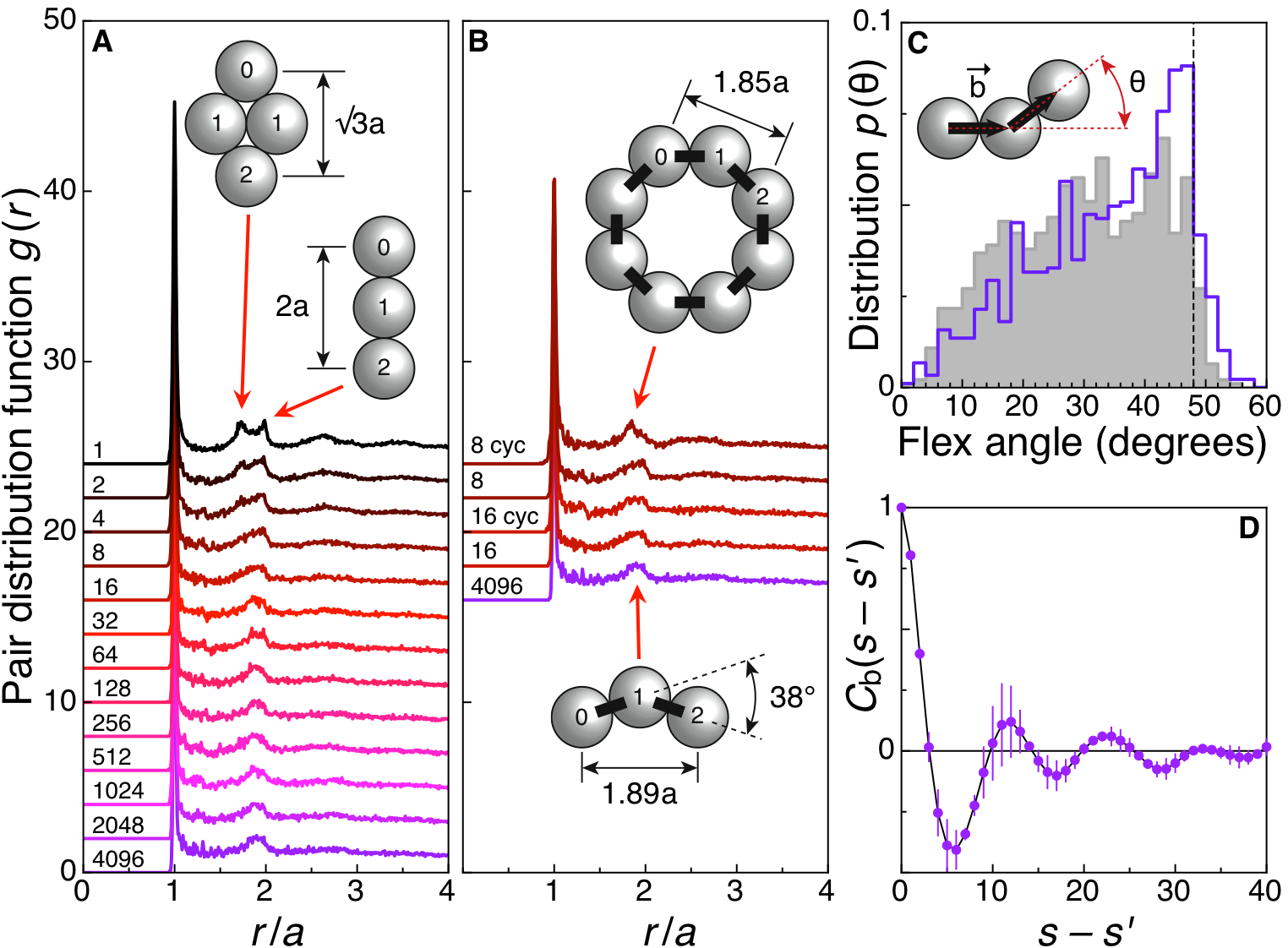}
\caption{The pair distribution function $g(r)$ for compacted packing of chains ($\xi=7.5$), curves are shifted for clarity: (a) Linear chains, from $M=1$ to $M=4096$. (b) Short cyclic (cyc) chains, compared with linear chains of the same length, and with a long linear chain. (c) The distribution of bond flex angles for long ($M=4096$, solid outline) and short ($M=8$, shaded fill) chain packings; dashed vertical line indicates $\theta_\mathrm{max}$. (d) The bond-orientational correlation $C_b(s-s')$ measured along a single chain, averaged over three long chain packings ($M=1024, 2048, 4096$).}
\label{g(r)}
\end{figure*}

Fig.~\ref{g(r)}a shows the pair distribution function $g(r)$ for packings of linear chains from $M=1$ to $M=4096$. The most notable change with $M$ is in the structure of the second peak, associated with second-nearest neighbors. For spheres ($M=1$), the second peak of $g(r)$ is split into two sub-peaks, a  well-known feature corresponding to two distinct particle arrangements \cite{SilbertLiuNagel2006}. The sub-peak at $r=2a$ corresponds to a linear trimer; for chains, this configuration naturally suggests three successive monomers along a backbone. The sub-peak at $r=\sqrt{3} a$ corresponds to a rhomboid cluster of 4 particles (Fig.~\ref{g(r)}a, inset); for chains, this requires contact between monomers without a shared bond.  As we move to long chains, the $r=\sqrt{3} a$ sub-peak quickly disappears, indicating a sharp suppression in the number of contacts between monomers that do not share a bond. For spheres, the mechanical stability of the packing is provided by pairwise contacts. For long chains, the suppression of pairwise contacts between monomers that do not share a bond leaves the backbone as the remaining source of rigidity. But how can floppy chains be arranged into a rigid structure?

\begin{figure*}[p]
\center
\includegraphics[width=0.75\textwidth]{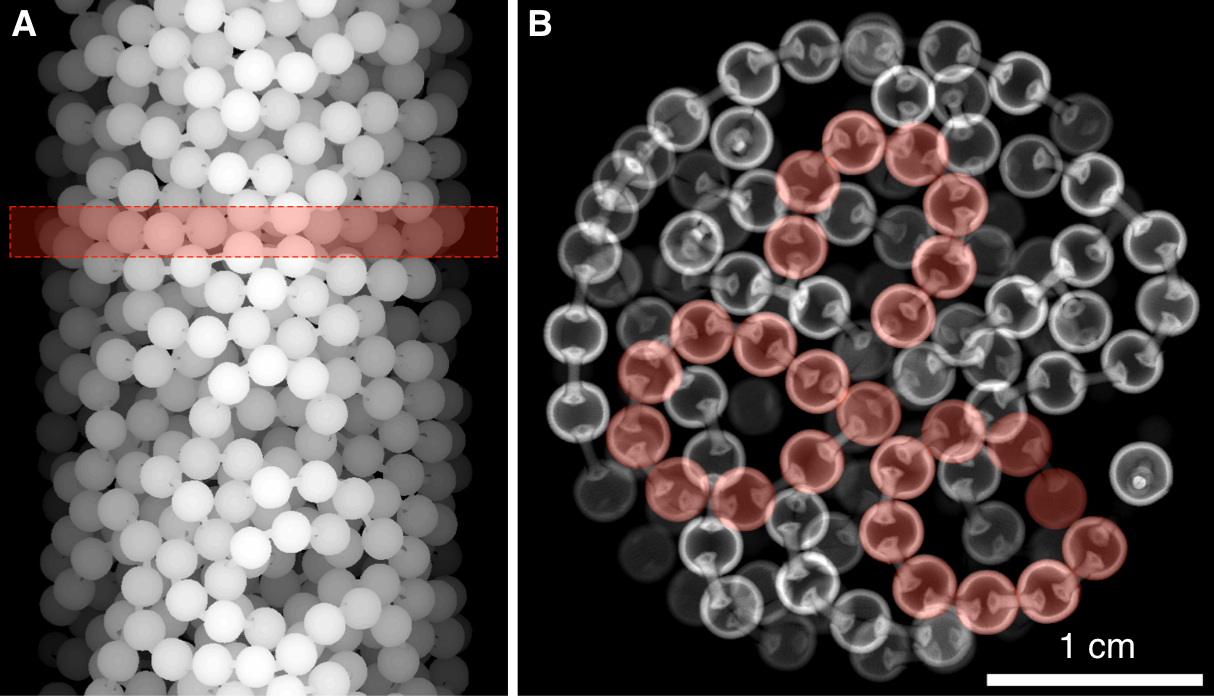}
\caption{Tomographic reconstruction of a packing that consists of a single long chain ($M=4096$): (a) The full reconstructed 3D image. (b) A small slice of the packing, indicated by the shaded box in (a), projected onto the horizontal plane.  The brighter a particle is, the closer it is to the center plane of the slice.  Highlighted particles are arranged in three near-minimal loops.}
\label{tomo}
\end{figure*}

Unlike the $r=\sqrt{3}a$ sub-peak, the sub-peak at $r=2a$ persists as $M$ increases: it broadens and shifts slightly to $r<2a$. This shift suggests a large number of bonds must be flexed; for $M=4096$, the peak center at $r=1.89a$ corresponds to a flex angle of $38^{\circ}$ (Fig.~\ref{g(r)}b, inset). Measured directly, the flex angle distribution $p(\theta)$ for long chains ($M=4096$) peaks strongly $\theta_\mathrm{max} = 48^\circ$; by contrast, $p(\theta)$ for short chains ($M=8$) is much flatter (Fig.~\ref{g(r)}c). We also compute the bond-orientational correlation $C_b(s-s')$ along a single chain
\begin{equation}
C_b(s-s') = \langle\vec{b}(s) \cdot \vec{b}(s')\rangle_s \ ;
\end{equation}
here $s, s'$ are monomer labels, and $\vec{b}(s)$ is the bond that links monomers $s$ and $s+1$.  Fig.~\ref{g(r)}d shows $C_b(s-s')$ for compacted long chains. It decays rapidly from unity, becomes maximum anti-correlated at $s-s' \simeq 6$, before returning to zero and executing small amplitude oscillations (period $\simeq 10.5a$). This indicates there is no long-range orientational order along the backbone; instead, the local chain conformations contain a prevalence of near-minimal, semi-rigid loops (recall $\xi = 7.5$) (Fig.~\ref{tomo}). For comparison, the $g(r)$ for $M=8$ cyclic chains, which are rigid loops, is very similar to that for long chains save for subtle differences: the second peak is sharper, and is located at $r=1.85a$ (rather than $1.89a$) corresponding to next-nearest vertices of an octagon. The $g(r)$ of the $M=16$ cyclic chain is nearly indistinguishable from that for long chains (Fig.~\ref{g(r)}b).

We suggest these strongly-flexed bonds and near-minimal loops are responsible for the rigidity of long-chain packings. Specifically, rigidity arises from the jamming of rigid elements (loops) that ``condensed'' out of a floppy object (the chain) upon compaction. Consider a pack of minimal loops, each with $\xi$ monomers; these will be completely rigid rings. A ring has 6 degree of freedom; for a pack of rings to be jammed, i.e., mechanically rigid, there must be on average 6 independent constraints per ring. These are provided by pairwise contacts between rings. On jamming, each ring is on average in contact with 12 other rings; on a per monomer basis, each monomer will have on average $12/ \xi$ contacts, or $z = 2+12/\xi$ nearest neighbors (coordination number). Compared with frictionless spheres, where the mean coordination at jamming is $z=6$, a jammed pack of rigid rings is much less coordinated and therefore less densely-packed. Since $z \sim 1/\xi$, the packing density will be even lower for stiffer chains (larger $\xi$).

The loops in compacted chains are nearly, but not exactly, minimal in size; they are semi-rigid and have more degrees of freedom than rigid rings. In addition, not all bonds are maximally flexed, and not all monomers belong on a loop. For linear chains, the chain ends, being bonded to only a single partner, are less constrained than monomers in the middle of a backbone. Each of these effects increases the number of constraints needed before a pack can jam, thereby enhancing the mean coordination and packing density at jamming. Therefore we expect: (1) long linear chains should pack less densely than short linear chains, but small, (nearly) rigid cyclic chains should pack the least densely of all; (2) stiff chains with large loop sizes should pack less densely than floppy chains with small $\xi$. Qualitatively, this is what we observed.

In the spirit of the Jamming Phase Diagram \cite{LiuNagel1998}, which links the glass transition of simple glass-formers with the jamming of grains, we suggest there is a similar connection between polymers and macroscopic chains. One well-studied aspect of the polymer glass transition is the variation of the glass transition temperature $T_\mathrm{g}$ on chain length and chain topology. For linear polymers, as $M\rightarrow \infty$, $T_\mathrm{g}$ quickly asymptotes to a constant value; whereas $T_\mathrm{g}$ decreases rapidly as $M\rightarrow 1$. This is commonly described by the Flory form
\begin{equation}\label{FloryForm}
T_\mathrm{g}(M) = T_{\mathrm{g}, \infty} - K/M\ ,
\end{equation}
where $K>0$ is polymer-specific parameter \cite{FoxFlory1949}. For cyclic polymers, in the long-chain limit, $T_\mathrm{g}$ is the same as that of long linear polymers of the same type; but as $M$ become very small, $T_\mathrm{g}$ increases \cite{Clarson&c1985}. As with ball chains, the behavior of linear and cyclic polymers follow opposite trends as $M$ becomes small.

\begin{figure}[p]
\center
\includegraphics[width=0.5\textwidth]{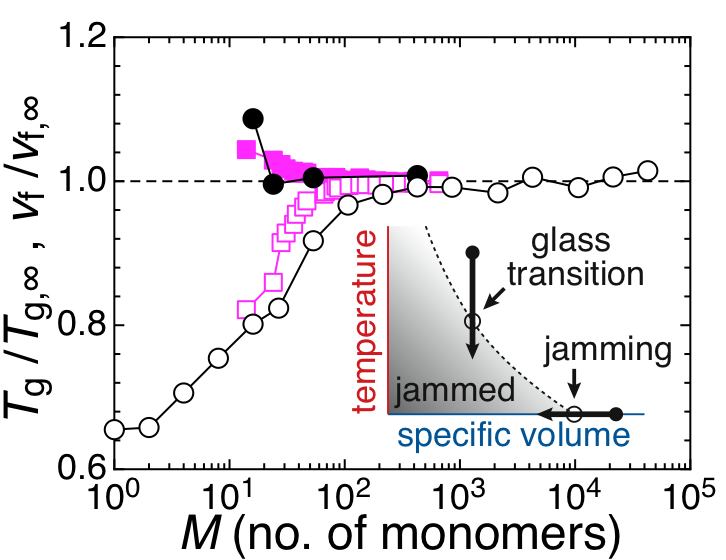}
\caption{The glass transition temperature $T_\mathrm{g}(M)$ of linear and cyclic polydimethylsiloxane (PDMS) (open and filled squares) compared with the specific packing volume $v_\mathrm{f}(M)\equiv1/\rho_\mathrm{f}(M)$ of linear and cyclic $\xi=11.2$ ball-chains (open and filled circles); both are normalized by their asymptotic $M\rightarrow \infty$ values $T_{\mathrm{g}, \infty}$ and $v_{\mathrm{f},\infty}$. PDMS data taken from \cite{Clarson&c1985}. Inset shows the $(T,v)$ plane of the Jamming Phase Diagram, with trajectories for glass transition and jamming.}
\label{comparison}
\end{figure}

The Jamming Phase Diagram suggests a way to directly compare glassy polymers with jammed chains. Here, temperature $T$ and specific packing volume $v\equiv1/\rho$ are orthogonal axes; jammed/glassy states occupy the high-density, low-temperature region of the phase diagram, the rest are occupied by fluid states. The glass transition is the transition from liquid to glass along the $T$-axis, while jamming is the transition from unjammed to jammed states along the $v$-axis at $T=0$ \cite{LiuNagel1998}. In this sense,  $T_\mathrm{g}(M)$ and  $v_\mathrm{f}(M) \equiv 1/\rho_\mathrm{f}(M)$ should be analogous quantities. When we compare $T_\mathrm{g}(M)/T_{\mathrm{g}, \infty}$ for linear and cyclic polydimethylsiloxane and $v_\mathrm{f}(M)/v_{\mathrm{f},\infty}$ for the packing of chains, the overall similarity between them is striking (Fig.~\ref{comparison}).

As shown earlier, in a packing of floppy linear chains, chain ends acts as non-interacting, density-enhancing defects. We can then write the specific packing volume $v_\mathrm{f}$ in the floppy regime as the weighted sum of $v_\mathrm{e}$, contributed by ends, and $v_\mathrm{b} > v_\mathrm{e}$, due to the bulk,
\begin{equation}
v_\mathrm{f}(M) = \left[1-\frac{2\ell}{M}\right]v_\mathrm{b} + \left[\frac{2\ell}{M}\right]v_\mathrm{e} = v_\mathrm{b} - \frac{2\ell(v_\mathrm{b} - v_\mathrm{e})}{M} \ ;
\end{equation}
here $\ell \sim \xi$ is the extent that end-influences propagate into the bulk of the chain.  Cast in this way, $v_\mathrm{f}(M)$ takes on a form identical to the Flory form (Eq. \ref{FloryForm}) for $T_\mathrm{g}(M)$. Fitted to appropriate regimes of the $\rho_\mathrm{f}(M)$ data, we find for $\xi = 7.5$, $v_\mathrm{b} = 2.3$, $\ell(v_\mathrm{b}-v_\mathrm{e}) = 2.6$; taking $\ell=\xi=7.5$, then $v_\mathrm{e} = 1.95$. For $\xi = 11.2$, the best fit gives $v_\mathrm{b} = 2.5$, and $\ell(v_\mathrm{b}-v_\mathrm{e})=6$; taking $\ell=\xi=11.2$ gives $v_\mathrm{e} = 1.96$.

Finally, chains with larger loop sizes pack less densely than chains with small $\xi$. The analogy of $v_\mathrm{f}(M)$ with $T_\mathrm{g}(M)$ suggests stiff polymers will have a higher $T_\mathrm{g}$ than those that are more flexible. Indeed, for vinyl polymers with rigid side-groups that restrict chain flexibility, $T_\mathrm{g}$ is higher for those with large side-groups than for those whose side-groups are small \cite{Mark&c1993}.

These observations do not prove the polymer glass transition must be due to the jamming of chains, especially at finite temperatures and with realistic interactions. Indeed, the role of temperature points to important distinctions between polymers and ball-chains: (1) the stiffness of a ball-chain is given solely by its construction, but a polymer becomes stiffer at lower $T$; (2) for the packing of macroscopic objects, thermal motion is irrelevant and the generalization of ``temperature'' is an open question \cite{EdwardsGrinev2002}. Therefore the purely geometric jamming of chains cannot be exactly analogous to the glass transition in polymers. Some aspects of temperature can perhaps be simulated by varying the tapping strength used to compact the packing \cite{Knight&c1995, Nowak1998}. Interaction can also be introduced, e.g. by making the chains slightly sticky with a thin coating of viscous oil. But even hard particles and no explicit temperature, the similarities between jammed ball-chains and glassy polymers are striking enough to suggest that the jamming idea captures much of the physics. Since it has long been a matter of debate whether glass transitions in polymers and in simple glass-formers are fundamentally similar, it is attractive to think jamming, whether of grains or chains, may provide a unifying connection. 

We have shown long floppy chains pack into a low-density structure whose rigidity is chiefly provided by the backbone. This can be understood as the jamming of semi-rigid loops that formed when the chains were compacted. By invoking an analogy between the specific packing volume and the glass transition temperature, we have shown the packing of chains parallels the polymer glass transition in important respects. If these two phenomena are indeed closely connected, as our data suggest, it would be a beautiful illustration of how molecular geometry and symmetry, independent of the specific microscopic interactions, can influence the structure of condensed matter. 


\begin{thebibliography}{99}
\bibitem{AsteWeaire2008}T. Aste, D. Weaire, \textit{The Pursuit of Perfect Packing} (Taylor \& Francis, Bristol, ed. 2, 2008).
\bibitem{Zallen1998}R. Zallen, \textit{The Physics of Amorphous Solids}, (Wiley-Interscience, New York, 1998).
\bibitem{Bernal1959}J. D. Bernal, \textit{Nature} \textbf{183}, 141 (1959).
\bibitem{Scott1962}G. D. Scott, \textit{Nature} \textbf{194}, 956 (1962)
\bibitem{Finney1970}J. L. Finney, \textit{Proc. R. Soc. Lond. Ser. A} \textbf{319}, 479 (1970).
\bibitem{Knight&c1995}J. B. Knight, C. G. Fandrich, C.-H. Lau, H. M. Jaeger, S. R. Nagel, \textit{Phys. Rev. E} \textbf{51}, 3957 (1995).
\bibitem{Villarruel&c2000}F. X.  Villarruel, B. E. Lauderdale, D. M. Mueth, H. M. Jaeger, \textit{Phys. Rev. E} \textbf{61}, 6914 (2000).
\bibitem{WilliamsPhillipse2003}S. R. Williams, A. P. Philipse, \textit{Phys. Rev. E} \textbf{67}, 051301 (2003). 
\bibitem{Donev&c2004}A. Donev \textit{et al.}, \textit{Science} \textbf{303}, 990 (2004).
\bibitem{Man&c2005}W. Man \textit{et al.}, \textit{Phys. Rev. Lett.} \textbf{94}, 198001 (2005).
\bibitem{Zeravcic&c2009}Z. Zeravcic, N. Xu, A. J. Liu, S. R. Nagel, W. van Saarloos, \textit{Europhys. Lett.} \textbf{87}, 26001 (2009).
\bibitem{Mailman2009}M. Mailman, C. F. Schreck, C. S. O'Hern, B. Chakraborty, \textit{Phys. Rev. Lett.} \textbf{102}, 255501 (2009).
\bibitem{Cahn1980}R. W. Cahn, \textit{Contemporary Physics} \textbf{21}, 43(1980).
\bibitem{JerniganFlory1969}R.L. Jernigan, P. J. Flory, \textit{J. Chem. Phys.}, \textbf{50}, 4178 (1969).
\bibitem{CarmesinKremer1988}I. Carmesin, K. Kremer, \textit{Macromolecules}, \textbf{21}, 2819 (1988).
\bibitem{SilbertLiuNagel2006}L. E. Silbert, A. J. Liu, S. R. Nagel, \textit{Phys. Rev. Lett.} \textbf{73}, 041304 (2006).
\bibitem{LiuNagel1998}A. J. Liu, S. R. Nagel, \textit{Nature} \textbf{396}, 21 (1998).
\bibitem{FoxFlory1949}T. G. Fox, P. J. Flory, \textit{J. Appl. Phys.} \textbf{21}, 581 (1949).
\bibitem{Clarson&c1985}S. J. Clarson, K. Dodgson, J. A. Semlyen, \textit{Polymer} \textbf{26}, 930 (1985).
\bibitem{Mark&c1993}A.  Eisenberg, in \textit{Physical Properties of Polymers}, J. E. Mark, Ed. (American Chemical Society, Washington DC, ed. 2, 1993).
\bibitem{EdwardsGrinev2002}S. F. Edwards, D. V. Grinev, \textit{Adv. Physics} \textbf{51}, 1669 (2002).
\bibitem{Nowak1998}E. R. Nowak, J. B. Knight, E. Ben-Naim, H. M. Jaeger, S. R. Nagel, \textit{Phys. Rev. E} \textbf{57}, 1971 (1998)
\bibitem{acknowledgements} We thank Peter Eng for his assistance, and Tom Witten for fruitful discussions. This work is supported by NSF MRSEC DMR-0820054 and DOE DE-FG02-03ER46088. GSECARS is supported by the NSF EAR-0622171 and DOE DE-FG02-94ER14466.   Use of the Advanced Photon Source is supported by the U. S. DOE, Office of Science, Office of Basic Energy Sciences, under Contract No. DE-AC02-06CH11357.
\end{thebibliography}
\end{document}